\documentclass[12pt,thmsa]{article}
\usepackage{sw20lart}

\input{tcilatex}
\begin{document}

\title{Are there local hidden variables models with time correlated detection
violating the Bell inequality?}
\author{Emilio Santos \\
Departamento de F\'{i}sica. Universidad de Cantabria. Santander. Spain}
\maketitle

\begin{abstract}
Explicit local hidden variables models are exhibited that assume a
correlation between detection events produced in the same detector at
different times. It is shown that some models give predictions closer to the
Bell limit than models without time correlation.
\end{abstract}

\section{Introduction}

Many experiments have been performed for the test of the Bell inequalities%
\cite{Genovese}. Amongst the experiments, those using photons are relevant
because they may allow closing the locality loophole more easily than those
using massive particles (like e. g. atoms\cite{Rowe}.) Until recently all
tests involving optical photons suffered from the ``detection loophole'',
that is the fact that entangled photons, when measured with low-efficiency
detectors, give results that may be reproduced by local hidden variables
models\cite{Santos}. However the progress in photon detectors, now available
with high efficiency and low noise, have allowed recent experiments free
from the detection loophole\cite{Giustina}, \cite{Kwiat}.

The loophole-free violation of a Bell inequality would be of paramount
importance. Indeed it would mean that no local hidden variables model (or
``local realistic theory'') exists compatible with empirical evidence\cite
{Bell}. Consequently it is relevant to search for any possible loophole in
the empirical tests. The purpose of this paper is to study whether some time
correlations between the production and/or detection of photons might give
rise to new loopholes. Such correlations may be more relevant for entangled
photons produce by parametric down conversion than for photons produced in
atomic cascades. In the latter it is likely that different atomic decays and
detections are uncorrelated. In contrast parametric down conversion produces
a beam where entangled photons appear spontaneously at random times, and
most probably with bunching due to the Bose character of photons. 

If there are no time correlations, for a given pair of photons emerging from
the source I will label $p_{a},p_{b}$, the single detection probabilities by
Alice and Bob, respectively, and $p_{ab}$ the coincidence detection
probability. Here we will consider the possibility that the detection of a
photon may be either enhanced or inhibited by a previous detection. In order
to make the study I will consider models where, after the detection by Alice
(Bob) of a photon belonging to the first ``photon pair'' (that I will name
``event'' for simplicity of writing) produced in the source, it is enhanced
or inhibited the detection probability of the Alice (Bob) photon belonging
to the second event. We assume similar correlations between the detections
of the third event and the fourth event, and so on excluding any other
correlation. In the experiment by Christensen et al.\cite{Kwiat} the
polarizer's settings are choosen at random, but only once every second. Many
photon pairs are produced during that time interval, so that time
correlations cannot be excluded.

To begin with we revisit a well known local hidden variables model. It has
one hidden variable, $\lambda $, with a homogeneous probability density, i.
e. 
\begin{equation}
\rho \left( \lambda \right) =\frac{1}{\pi };\lambda \in \left[ 0,\pi \right]
,  \label{1}
\end{equation}
and the detection probability by Alice, given $\lambda $ and $\alpha ,$ is
assumed to be 
\begin{eqnarray}
P(\lambda ,\alpha ) &=&\frac{1}{6}\left[ 1+\sqrt{2}\cos \left( \lambda
-\alpha \right) \right] ^{2}  \nonumber \\
&\equiv &\frac{1}{3}\left[ 1+\sqrt{2}\cos \left( 2\lambda -2\alpha \right) +%
\frac{1}{2}\cos \left( 4\lambda -4\alpha \right) \right] ,  \label{2}
\end{eqnarray}
which may be checked to fulfil 
\[
0<P(\lambda ,\alpha )<1,
\]
as it should. Similarly for Bob 
\begin{equation}
P(\lambda ,\beta )=\frac{1}{3}\left[ 1+\sqrt{2}\cos \left( 2\lambda +2\beta
\right) +\frac{1}{2}\cos \left( 4\lambda +4\beta \right) \right] .  \label{3}
\end{equation}
Hence we get the following coincidence and single probabilities 
\begin{eqnarray}
p_{ab}\left( \phi \right)  &=&\frac{1}{9}\left[ 1+\cos \left( 2\phi \right) +%
\frac{1}{8}\cos \left( 4\phi \right) \right] ,  \nonumber \\
\phi  &\equiv &\alpha -\beta ,p_{a}\left( \alpha \right) =p_{b}\left( \beta
\right) =1/3,  \label{A}
\end{eqnarray}
where $\alpha $ and $\beta $ are the angles of the polarizer's settings
with, say, the vertical. The Bell inequality in the form of Clauser and
Horne may be written 
\begin{equation}
B\equiv \frac{3p_{ab}\left( \phi \right) -p_{ab}(3\phi )}{p_{a}+p_{b}}\leq 1,
\label{CH}
\end{equation}
and it is fulfilled by the local model predictions for any $\phi ,$ as it
should. In particular for the usually measured angles $\phi =\pi /8,3\phi
=3\pi /8$ we get 
\begin{equation}
B=\frac{3\times \frac{1}{9}\left( 1+\frac{\sqrt{2}}{2}\right) -\frac{1}{9}(1-%
\frac{\sqrt{2}}{2})}{2\times \frac{1}{3}}=\frac{3\times 0.190-0.032}{0.667}%
=0.805<1.  \label{B}
\end{equation}

The predictions of this model are close to the quantum predictions for
experiments with detectors having efficiency about 67\%. In particular the
value of the parameter B, eq.$\left( \ref{B}\right) $ exactly reproduces the
quantum prediction with detector efficiency 2/3. This suggests that a
similar model for experiments involving photons not maximally entangled,
similar to the recent detection-loophole-free ones\cite{Giustina}, \cite
{Kwiat}, might provide values much closer to the Bell limit (that is the
corresponding parameter similar to B closer to unity). However such models
would be more involved than the one given by eqs.$\left( \ref{1}\right) $ to 
$\left( \ref{3}\right) .$

\section{Model with inhibited detection}

Now we consider a modification of the model by assuming that Alice's
detection of her photon in the second event is inhibited if she detected her
photon in the first event, and similar for Bob. Thus the probability of two
detections from two events will be zero, for a single detection from the
first event (and no detection in the second event) will be $p_{a}$, the
probability of a single detection from the second event will be $\left(
1-p_{a}\right) p_{a}$. Finally the probability of zero detections will be $%
\left( 1-p_{a}\right) ^{2}.$ Therefore the mean probability of detection per
event will be $\left( 2-p_{a}\right) p_{a}$%
\[
p_{a}^{\prime }=\frac{1}{2}\left( 2-p_{a}\right) p_{a}=p_{a}-\frac{1}{2}%
p_{a}^{2}. 
\]
The same is true for Bob.

Now we shall study coincidences. In the first event the probability of one
or zero coincidences will be $p_{ab}$ or $1-p_{ab},$ respectively, and the
probability of a coincidence in the second event is zero if either Alice or
Bob or both had one detection in the first event. Thus the probability of
having one coincidence in the second event will be $\left(
1-2p_{a}+p_{ab}\right) p_{ab}.$ (Remember that we assume $p_{a}=p_{b}).$ The
mean probability of coincidence per event will be 
\[
p_{ab}^{\prime }=\frac{1}{2}\left[ p_{ab}+\left( 1-2p_{a}+p_{ab}\right)
p_{ab}\right] =p_{ab}-p_{a}p_{ab}+\frac{1}{2}p_{ab}^{2}. 
\]
Hence the parameter eq.$\left( \ref{CH}\right) $ becomes 
\begin{eqnarray*}
B^{\prime } &=&\frac{3p_{ab}^{\prime }\left( \phi \right) -p_{ab}^{\prime
}\left( 3\phi \right) }{2p_{a}^{\prime }}=\frac{\left( 1-p_{a}\right) \left[
3p_{ab}\left( \phi \right) -p_{ab}\left( 3\phi \right) \right] +\frac{3}{2}%
p_{ab}\left( \phi \right) ^{2}-\frac{1}{2}p_{ab}\left( 3\phi \right) ^{2}}{%
2\left( p_{a}-\frac{1}{2}p_{a}^{2}\right) } \\
&=&\frac{1-p_{a}}{1-\frac{1}{2}p_{a}}B+\frac{3p_{ab}\left( \phi \right)
^{2}-p_{ab}\left( 3\phi \right) ^{2}}{2p_{a}\left( 2-p_{a}\right) }.
\end{eqnarray*}
Inserting the values of eqs.$\left( \ref{A}\right) $ and $\left( \ref{B}%
\right) $ we get 
\[
B^{\prime }=0.805\times \frac{2/3}{5/6}+\frac{3\times 0.190^{2}-0.032^{2}}{%
10/9}=0.644+0.096=0.740. 
\]

\section{Model with enhanced detection}

In this case, if Alice detects one photon from the first event, she will
detect another one in the second event with certainty. The probability of
this situation is $p_{a}$. If she does not detect in the first event
(probability $1-p_{a}),$ she may detect in the second event. The total
probability of this is $p_{a}\left( 1-p_{a}\right) .$ In summary the mean
single probability per event is 
\[
p_{a}^{\prime \prime }=\frac{3}{2}p_{a}-\frac{1}{2}p_{a}^{2}. 
\]
If there is a coincidence count in the first event (probability $p_{ab})$
there will be another in the second event with certainty. If there is a
single detection by Alice in the first pair (probability $p_{a}-p_{ab})$
there will be another one in the second event. The probability that also Bob
detects in the second event will be $p_{a},$ so the probability of one
single count in the first event and one coincidence in the second event will
be $2\left( p_{a}-p_{ab}\right) p_{a}.$ If there is no detection in the
first event (probability $1-2p_{a}+p_{ab})$ the probability of coincidence
in the second event will be $p_{ab}.$ In summary, the mean probability of
coincidence count per event will be 
\begin{eqnarray*}
p_{ab}^{\prime \prime } &=&\frac{1}{2}\left[ 2p_{ab}+2\left(
p_{a}-p_{ab}\right) p_{a}+\left( 1-2p_{a}+p_{ab}\right) p_{ab}\right] \\
&=&\frac{3}{2}p_{ab}-2p_{a}p_{ab}+\frac{1}{2}p_{ab}^{2}+p_{a}^{2}.
\end{eqnarray*}
The parameter eq.$\left( \ref{CH}\right) $ becomes 
\begin{eqnarray*}
B^{\prime \prime } &=&\frac{\left( \frac{3}{2}-2p_{a}\right) \left[
3p_{ab}\left( \phi \right) -p_{ab}\left( 3\phi \right) \right] }{(\frac{3}{2}%
-\frac{1}{2}p_{a})2p_{a}}+\frac{3p_{ab}^{2}\left( \phi \right)
-p_{ab}^{2}\left( 3\phi \right) }{(3-p_{a})2p_{a}}+\frac{2p_{a}}{3-p_{a}} \\
&=&\frac{3-4p_{a}}{3-p_{a}}B+\frac{3p_{ab}\left( \phi \right)
^{2}-p_{ab}\left( 3\phi \right) ^{2}}{2p_{a}\left( 3-p_{a}\right) }+\frac{%
2p_{a}}{3-p_{a}}.
\end{eqnarray*}
Putting the values of eqs.$\left( \ref{A}\right) $ and $\left( \ref{B}%
\right) $ we get 
\[
B^{\prime \prime }=\frac{5}{8}\times 0.805+\frac{3\times 0.190^{2}-0.032^{2}%
}{16/9}+\frac{1}{4}=0.503+0.060+0.25=0.813. 
\]
The enhancement in detection increases but slightly the parameter $B$.

\section{Conclusion}

The time correlation in detection is able to increase the value of the Bell
parameter eq.$\left( \ref{CH}\right) $ in the model eqs.$\left( \ref{1}%
\right) $ to $\left( \ref{3}\right) $. The effect is too small to allow a
violation of the Bell inequality. However a small increase might be enough
in models of the experiments involving photons not maximally entanbled,
similar to the recent detection-loophole-free ones\cite{Giustina}, \cite
{Kwiat}. Indeed in this case a model without time correlations could provide
values close to the Bell limit, as mentioned at the end of Section 2. This
possibility will be studied elsewhere.

\end{document}